\documentclass[aps,prl,twocolumn,superscriptaddress,showpacs]{revtex4-1}

\usepackage{graphicx}
\usepackage{longtable}
\usepackage[ansinew]{inputenc}
\usepackage{color}

\newcommand{\ang}{\ensuremath{\text{\AA}}}

\newcommand{\bi}{\mathbf{i}}
\newcommand{\bj}{\mathbf{j}}
\newcommand{\bk}{\mathbf{k}}
\newcommand{\bq}{\mathbf{q}}
\newcommand{\up}{\uparrow}
\newcommand{\dn}{\downarrow}
\newcommand{\ket}[1]{\ensuremath{|#1\rangle}}
\newcommand{\bra}[1]{\ensuremath{\langle #1|}}

\newcommand{\beq}{\begin{eqnarray}}
\newcommand{\eeq}{\end{eqnarray}}

\begin{document}

\title{Strength of effective Coulomb interactions in graphene and graphite}

\author{T. O. Wehling}
\affiliation{1.\ Institut für Theoretische Physik, Universität Hamburg,
D-20355 Hamburg, Germany}
\author{E. \c{S}a\c{s}{\i}o\u{g}lu}
\author{C. Friedrich}
\affiliation{Peter Grünberg Institut and
Institute for Advanced Simulation, Forschungszentrum Jülich and JARA,
D-52425 Jülich, Germany}
\author{A. I. Lichtenstein}
\affiliation{1.\ Institut für Theoretische Physik, Universität Hamburg,
D-20355 Hamburg, Germany}
\author{M. I. Katsnelson}
\affiliation{Radboud University Nijmegen, Institute for Molecules
and Materials, NL-6525 AJ Nijmegen, The Netherlands}
\author{S. Bl\"{u}gel$^2$}

\begin{abstract}
To obtain an effective many-body model of graphene and
related materials from first principles we calculate the partially
screened frequency dependent Coulomb interaction. 
In graphene, the
effective on-site (Hubbard) interaction is $U_{00}=9.3$\,eV in close vicinity to
the critical value separating conducting graphene from an insulating phase emphasizing the importance of non-local Coulomb terms.
The nearest-neighbor Coulomb interaction strength is computed to $U_{01}=5.5$\,eV.
In the long wavelength limit, we find the
effective background dielectric constant of graphite to be
$\epsilon=2.5$ in very good agreement with experiment.
\end{abstract}


\maketitle
The role of Coulomb interactions in graphene and related materials
poses a long standing problem: Experiments reported ferromagnetic
ordering in nanographene
\cite{Shibayama_PRL00,*Enoki_SolidStateCommun09}, in disordered
graphite samples \cite{Esquinazi_PRB02,*Esquinazi_PRL03} and at
grain boundaries in highly oriented pyrolytic graphite (HOPG)
\cite{Flipse_NatPhys09}. Ferromagnetism in pristine graphene,
however, has been excluded experimentally for temperatures down to
2K \cite{Grigorieva_PRL10}. Theoretically, the possibility of
magnetism in defect free graphene has been predicted: An
antiferromagnetic insulating ground state has been obtained for
the local Coulomb interactions exceeding a critical value $U_{\rm
AF}\gtrsim (4.5 \pm 0.5)t$ in Quantum Monte Carlo (QMC) calculations
\cite{Sorella_92,Martelo_ZPhysB97,Pavia_PRB05} and $U_{\rm
AF}\gtrsim 2.2t$ in Hartree-Fock theory
\cite{Sorella_92,Martelo_ZPhysB97}, where $t\approx 2.8\,$eV is the nearest neighbor hopping parameter. A gapped spin-liquid has been predicted for on-site repulsion between $U_{\rm sl}=3.5t$ and $U_{\rm
AF}$ \cite{Muramatsu_Nature10}. Sizable non-local Coulomb interactions can make the phase diagram even richer and lead to a competition between spin- and charge-density-wave phases \cite{herbut,Honerkamp_CDW_SDW_PRL08} or topologically non-trivial phases \cite{Honerkamp_TI_PRL08}. Doping of graphene might trigger further instabilities \cite{Peres_PRB2004,Baskaran_PRB10}.
In pristine graphene, the
Coulomb interaction remains long ranged and it is controversial
whether this might lead to strongly correlated electronic phases
like an insulator \cite{herbut,Joaquin_PRL09} or whether graphene
is rather weakly correlated. The local part of Coulomb interaction
is also crucial for the theory of defect-induced magnetism in graphene
\cite{vozmediano}.

The central issue in this discussion is the effective strength of
the Coulomb interaction acting on the carbon $p_z$-electrons,
which has only been estimated very roughly up to now
\cite{RMP_AHC2009}: The bare on-site Coulomb interaction in
benzene obtained from atomic carbon $p_z$ orbitals was estimated
to be $16.9$\,eV \cite{Parr_JCP50}. For polyacetylene, an analysis
of optical modulation spectroscopy experiments within weak
coupling perturbation theory yielded an effective on-site Coulomb repulsion of
$10$\,eV \cite{Vardeny_PRL85,Baeriswyl_PRL86}. However, in this
regime weak coupling perturbation theory might be inapplicable.
For the long wavelength limit, reflectance measurements of
graphite \cite{Taft_PR65} yielded a dielectric constant of
$\epsilon=2.4$ due to screening by the high energy $\sigma$-bands.
This would correspond to an effective fine structure constant of
$\alpha=\frac{e^2}{\epsilon\hbar v_F}\approx 0.9$ for bulk graphite, where $\hbar v_F\approx 5.8$\,eV$\ang$ is the Fermi velocity  \cite{RMP_AHC2009}. 
For graphene, recent
inelastic x-ray scattering experiments \cite{Reed_Science10}
suggest a fully screened dielectric constant of
$\epsilon\approx 15$ corresponding to a fine structure constant of
$\alpha=0.14$. At the same time, first-principles $GW$ calculations
\cite{SK2010} give $\epsilon\approx 4$, in agreement with the
predictions of a simple Dirac model \cite{RMP_AHC2009}. Recent experimental data on charge density dependence of the
Fermi velocity \cite{NovoselovGeimGuinea_11} seem to be in agreement, rather, with the second
value. So, up to
now the strength of Coulomb interactions in graphene related
materials has remained unclear and controversial
--- both theoretically and experimentally (for a review of correlation
effects in graphene, see Ref. \onlinecite{RMP_2011}).

\begin{table}%
\begin{tabular}{|l|l|l||l|l|}
\hline
 & \multicolumn{2}{|c||}{graphene} & \multicolumn{2}{c|}{graphite}\\
 \hline
 & bare& cRPA& bare& cRPA\\
\hline
$U^{A/B}_{00}$ (eV)& 17.0 & 9.3 & 17.5, 17.7 & 8.0, 8.1 \\
$U_{01}$\hfill (eV) & 8.5 & 5.5 & 8.6 & 3.9\\
$U^{A/B}_{02}$ (eV) & 5.4 & 4.1 & 5.4, 5.4 & 2.4, 2.4\\
$U_{03}$\hfill (eV) & 4.7 & 3.6 & 4.7& 1.9 \\
\hline
\end{tabular}
\caption{On-site ($U^{A}_{00}$, $U^{B}_{00}$), nearest-neighbor
($U_{01}$), next-nearest-neighbor ($U^{A}_{02}$, $U^{B}_{02}$), and third-nearest-neighbor ($U_{03}$) (intra-layer) Coulomb interaction parameters for
freestanding graphene and graphite. In graphene
$U^A_{00}=U^B_{00}$ and $U^A_{02}=U^B_{02}$ due to the sublattice symmetry. The bare and
partially screened (cRPA) parameters are given. The cRPA
parameters should be used in the effective Hamiltonian
(\ref{eqn:Hubbard_graphene}).} \label{tab:Umat_graphene_graphite}
\end{table}
In this letter, we determine the Coulomb interaction strength in
graphene and graphite within the constrained random phase
approximation (cRPA)
\cite{Aryasetiawan_PRB04,*Aryasetiawan_PRB06}. We obtain
\textit{ab initio} effective Coulomb interaction parameters that
should be used in a generalized Hubbard model of graphene or
graphite (see cRPA values in table \ref{tab:Umat_graphene_graphite}). We find that the on-site interactions in free standing graphene are weaker than $U_{\rm AF}$ but close to the transition to the insulating spin liquid phase at $U_{\rm sl}=3.5t\approx 9.8\,$eV. Our calculations stress the importance of non-local Coulomb interactions in graphene. They put graphene in close proximity to two quantum phase transition lines and at the same time are possibly crucial for stabilizing a conducting state of freely suspended graphene. In the
long wavelength limit, we find bulk graphite having an effective
background dielectric constant $\epsilon\approx 2.5$, in
agreement with the experiments from Ref. \onlinecite{Taft_PR65}.
For graphene in the long-wavelength limit $\epsilon$ is just one, as it
should be for any two-dimensional system as will be explained below.

We start with constructing a generalized Hubbard model for the
graphene $\pi$-bands,
\begin{eqnarray}
 \hat{H_0}&=&-t\sum_{<\bi,\bj>,\sigma}c^\dagger_{\bi,\sigma} c_{\bj,\sigma}-t'\sum_{\ll \bi,\bj\gg,\sigma}c^\dagger_{\bi,\sigma} c_{\bj,\sigma}\nonumber\\
&&+U_{00}\sum_{\bi}n_{\bi,\up}n_{\bi,\dn}+\frac{1}{2}\sum_{\bi\neq\bj,\sigma,\sigma'}U_{\bi\bj}n_{\bi,\sigma}n_{\bj,\sigma '},
\label{eqn:Hubbard_graphene}
\end{eqnarray}
where $c_{\bi,\sigma}$ annihilates an electron with spin
$\sigma\in\{\up,\dn\}$ at site $\bi$ and
$n_{\bi,\sigma}=c^\dagger_{\bi,\sigma} c_{\bi,\sigma}$. The index
$\bi=(i,A/B)$ labels the sublattice (A,B) and the unit cell
centered at position $R_i$, $U_{\bi\bj}$ are the Coulomb interaction
parameters. The nearest neighbor hopping is known to be $t\approx
2.8$\,eV \cite{Reich:2002,RMP_AHC2009} and the next-to-nearest
neighbor hopping $t'$ depends on details of how the tight-binding
parameters are determined: $0.02t \lesssim t'\lesssim 0.2t$.

To obtain all parameters entering the Hamiltonian
(\ref{eqn:Hubbard_graphene}) from first principles, we performed
density functional theory (DFT) and cRPA calculations. The DFT
calculations are carried out with the FLEUR code \cite{FLEUR}
using a generalized gradient approximation \cite{PBE_96} for the exchange-correlation energy functional. We use a linear momentum cutoff of 
$G_{\rm max}=4.5\,\text{bohr}^{-1}$ for the plane waves and an angular momentum cutoff of $l_{\rm
max}=6$ in the muffin-tin spheres. The partially screened Coulomb
matrix elements are calculated in the cRPA with the SPEX code
\cite{Friedrich_CPhysCommun09,*Friedrich_PRB10,Sasioglu_PRB10,*Sasioglu_PRB11}
using the mixed product basis
\cite{Aryasetiawan_PRB94,Kotani_SolidStateCommun02,Friedrich_CPhysCommun09}
with cutoff values $G'_{\rm max}=4\,\text{bohr}^{-1}$ and $L_{\rm max}=4$. 


The Hamiltonian (\ref{eqn:Hubbard_graphene}) describes a system of C-$p_z$ electrons that
interact via the effective interaction $U_{\bi\bj}$, which incorporates the
screening effects of all other electrons not contained in the Hamiltonian (\ref{eqn:Hubbard_graphene}).
The cRPA approach offers an efficient way to calculate this interaction \cite{Aryasetiawan_PRB04,*Aryasetiawan_PRB06},
as the screening channels are individually accessible. The
two-dimensional symmetry of graphene clearly separates the C-$p_z$ from
other bands and, thus, enables an unequivocal elimination of the C-$p_z$
screening from the full RPA polarization function. Apart from the on-site
term the resulting effective interaction yields the
off-site, intra-orbital, and inter-orbital terms as well as their
frequency dependence.

The fully screened long wavelength dielectric constants reported in Refs. \cite{Reed_Science10,SK2010,NovoselovGeimGuinea_11} are different from the partially screened cRPA dielectric constants obtained, here, in that the former include also contributions to screening due to transitions between the graphene $\pi$ bands. Hence, using the dielectric constants from Refs. \cite{Reed_Science10,SK2010,NovoselovGeimGuinea_11} in a generalized Hubbard model like Eq. (\ref{eqn:Hubbard_graphene}) or in the context of investigations like Refs. \cite{herbut,Joaquin_PRL09} would lead to double counting of screening terms arising from the $\pi$ electrons. 

We ensure the accuracy of the model parameters being derived by carefully checking
their dependence on the calculation procedure (the type of
Wannier construction being used to define the C-$p_z$ orbitals) and convergence issues (Brillouin zone sampling and finite supercell height $h$) as we explain in the online supporting material \cite{EPAPS}. We find that Wannier functions directly from the C-$p_z$ projections \cite {Pickett_PRL02} and $16\times 16\times 1$ k-meshes for the BZ integration yield accurate Coulomb interaction parameters.

For graphene at its equilibrium lattice constant of $a_0=2.47\,\ang$, we obtain the Coulomb interaction parameters given in table \ref{tab:Umat_graphene_graphite}. The
on-site Coulomb repulsion $U^{A/B}_{00}\approx 3.3t$ is
below $U_{\rm AF}\approx (4.5 \pm 0.5)t$
\cite{Sorella_92,Martelo_ZPhysB97,Pavia_PRB05} but very close to the critical value of $U_{\rm sl}=3.5t$ separating the zero gap phase from a gapped spin liquid one \cite{Muramatsu_Nature10}. Comparing to the phase diagram reported in Ref. \cite{Honerkamp_CDW_SDW_PRL08} our results show that the nearest-neighbor Coulomb interaction of $U_{01}\approx 2.0 t$ taken together with the local Coulomb interaction puts graphene in close proximity to, both, a spin-density wave and a charge density wave transition line.
\begin{figure}%
\includegraphics[width=0.98\columnwidth]{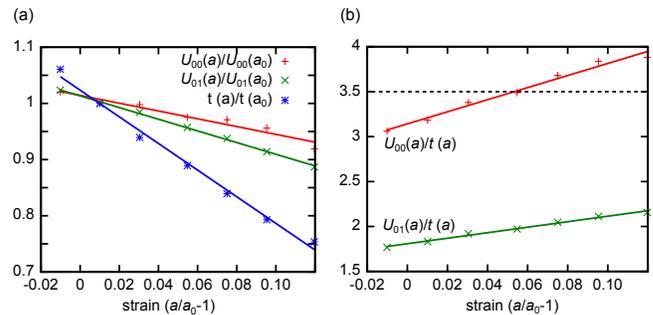}%
\caption{(Color online) Effect of lattice expansion on the strength of Coulomb interactions obtained with $h=21.2\,\ang$. (a) On-site $U_{00}(a)/U_{00}(a_0)$ and nearest neighbor $U_{01} (a)/U_{01}(a_0)$ Coulomb interaction as well as nearest-neighbor hopping $t(a)/t(a_0)$ as function of isotropic strain $(1-a/a_0)$. $a$ is the lattice constant. The parameters are given relative to their values at the equilibrium lattice constant $a_0=2.47\,\ang$. The solid lines are linear fits serving as guide to the eye. (b) Ratios of on-site $U_{00}(a)/t(a)$ and nearest-neighbor $U_{01}(a)/t(a)$ Coulomb interaction to the nearest-neighbor hopping. The dashed line indicates the phase boundary at $U_{\rm sl}/t=3.5$ separating the zero gap phase from a gapped spin liquid one \cite{Muramatsu_Nature10}.}%
\label{fig:U_t_lp change}%
\end{figure}

The ratio of the kinetic energy given by $t$ to the Coulomb interaction can, e.g., be changed by applying strain. Upon expanding the graphene lattice the nearest neighbor hopping decreases faster than the Coulomb interaction parameters (Fig. \ref{fig:U_t_lp change} a).
An expansion of the lattice by a few percent leads to $U_{00}(a)/t(a)>3.5$, i.e. an increase of the ratio of local Coulomb interactions to the kinetic energy beyond the critical value of $U_{\rm sl}/t=3.5$. In this situation, the non-local Coulomb interaction effects can be crucial. It remains to be seen to which extent the long range non-local Coulomb interaction screens the on-site repulsion \cite{Chitra_PRL00,*Ping_PRB02} and stabilizes the semimetallic phase or whether non-local Coulomb terms drive the system towards other strongly correlated possibly topologically non-trivial electronic phases as suggested in Refs. \cite{Honerkamp_CDW_SDW_PRL08,Honerkamp_TI_PRL08}.

We now consider the Coulomb interaction in graphite and compare to
graphene. In graphite, the two sublattices are not equivalent. We
define the atoms of sublattice A be to directly above each other
in adjacent layers and sublattice B as the atoms above hollow
sites of the layer beneath. As table
\ref{tab:Umat_graphene_graphite} shows, the on-site interaction in
graphene and graphite is qualitatively similar with very little
difference between the two graphite sublattices. The ratio of bare
to cRPA nearest-neighbor Coulomb interaction is
$U^{\text{bare}}_{01}/U^{\text{cRPA}}_{01}=1.6$ in graphene as
compared to 2.2 in graphite. The non-local screening by the
$\sigma$-bands is considerably more effective in graphite
than in graphene.

\begin{figure}%
\includegraphics[width=0.98\columnwidth]{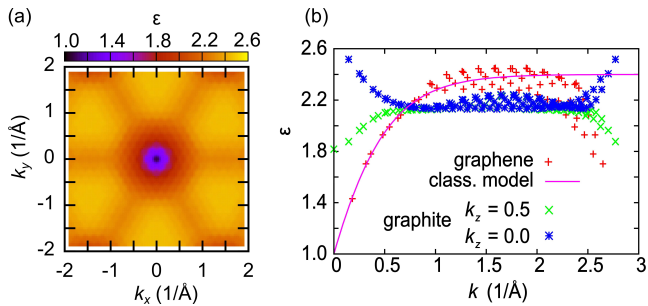}
\caption{(Color online) Static cRPA dielectric functions $\epsilon(\bk)$ of graphene and graphite as function of (in-plane) momentum transfer $\bk=(k_x,k_y)$. (a) Color coded (grayscale) $\epsilon(\bk)$ for graphene. The most pronounced effect is the decrease of $\epsilon(\bk)$ for $\bk\to 0$. There is a small directional modulation at intermediate momentum transfer $k=|\bk|\gtrsim 1\,\ang^{-1}$. (b) cRPA dielectric functions $\epsilon(k)$ of graphene and graphite as function of $k=|\bk|$. Eq. (\ref{eq:class_model_graphene}) fits well the background dielectric screening for freestanding graphene in the limit of $k\to 0$. 
For graphite, two values of the perpendicular momentum transfer are considered: $k_z=0$ and $k_z=0.5 (2\pi/c)$ with $c=3.3\,\ang$ being the graphite interlayer spacing.}%
\label{fig:epsilon_k}%
\end{figure}

This trend manifests clearly in the long wavelength limit
as can be seen from the Coulomb interaction in reciprocal space. 
To this end, we consider the Coulomb interaction matrix elements
in terms of the Bloch transformed C-$p_z$-Wannier functions,
$\ket{w_{n\bk}}$. We calculate the ratio of bare to cRPA screened
interaction \footnote{Here, $\epsilon(\bk)$ is defined as the ratio of the bare and partially
screened electron-electron interaction potentials with momentum transfer
$\bk$, i.e., $\bq_1\to\bq_1+\bk$ and $\bq_2+\bk\to\bq_2$, averaged over the momenta $\bq_1$ and $\bq_2$.
The Wannier function index $n$ is chosen to correspond to atoms in sublattice A.}
\begin{equation}
\epsilon(\bk)=\frac{\bra{w_{n\bq_1}w_{n\bq_2+\bk}}W^{\rm bare}\ket{w_{n\bq_1+\bk}w_{n\bq_2}}}{\bra{w_{n\bq_1}w_{n\bq_2+\bk}}W^{\rm cRPA}\ket{w_{n\bq_1+\bk}w_{n\bq_2}}}.
\end{equation}
For graphene, our cRPA calculations (Fig. \ref{fig:epsilon_k})
yield $\epsilon(\bk)\approx 2.4$ for intermediate momentum
transfer, $k=|\bk|\gtrsim 1\,\ang^{-1}$, and $\epsilon(\bk)\to 1$
for $k\to 0$. The screening due to high energy states in graphene
becomes essentially negligible in the long wavelength limit \footnote{The long wavelength behavior of $\epsilon(k)$ determines the screening of the long range tails of the Coulomb interaction. $\epsilon(k)\to 1$ for $k\to 0$ corresponds to an unscreened $1/r$ tail of the Coulomb interaction.}. This
is fundamentally different for graphite where
$\epsilon(k_{||})\approx 2$ almost independently of the momentum
transfer and $\epsilon(k=0)\approx 2.5$. Hence, graphite should be less correlated than graphene.

In the long wavelength limit, the simplest model to address
screening by high-energy bands in freestanding graphene is to
consider a film of thickness $d$ and dielectric constant
$\epsilon_1$. Transferring Ref. \cite{Emelyanenko_JPhys08} to the geometry at hand \footnote{Here, we consider a point charge in the middle of the film and evaluate the Coulomb potential in the middle of the film. Using the conventions and nomenclature of Ref. \cite{Emelyanenko_JPhys08} our situation corresponds to $z=0$, $d=h/2$, $\kappa=0$, $\epsilon_2=\epsilon_3=1$ and $\beta_{13}=\beta'_{12}=(\epsilon_1-1)/(\epsilon_1+1)$. Then, Eq. (4) of Ref. \cite{Emelyanenko_JPhys08} leads after division by the bare interaction ($q/\lambda$) to $\epsilon(k=\lambda)$ as in our Eq. (\ref{eq:class_model_graphene}). } we obtain
\begin{eqnarray}
\epsilon_1^{-1}(\bk)&=&\frac{1}{\epsilon_1}\cdot\frac{\epsilon_1+1+(\epsilon_1-1)e^{-kd}}{\epsilon_1+1-(\epsilon_1-1)e^{-kd}}
\label{eq:class_model_graphene}\\
&\stackrel{k\to 0}{\longrightarrow}&1+kd\left(\frac{1}{2\epsilon_1}-\epsilon_1+1/2\right).
\end{eqnarray}
Our cRPA calculations confirm this expectation (see Fig.
\ref{fig:epsilon_k}). Eq. (\ref{eq:class_model_graphene}) turns
out to describe the partially screened Coulomb interaction well
for $k=|\bk|<1\,\ang^{-1}$ with $d=2.8\ang$ and $\epsilon_1=2.4$
proving the applicability of this classical model at long
wavelengths.

\begin{figure}%
\includegraphics[width=0.75\columnwidth]{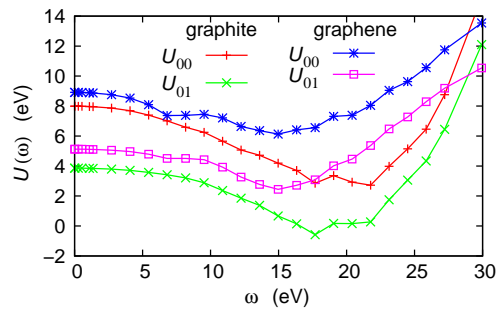}%
\caption{(Color online) Frequency dependence of the on-site and nearest-neighbor interaction obtained from cRPA for graphene ($h=21.2\,\ang$) and graphite. For graphite $U_{00}(\omega)=U_{00}^A(\omega)$ is shown, which is virtually the same as $U_{00}^B(\omega)$. $|U_{00}^A(\omega)-U_{00}^B(\omega)|<0.15$\,eV for $\omega< 20$\,eV.}%
\label{fig:U_omega}%
\end{figure}

Integrating out the graphene $\sigma$-bands and other high energy
states leads to frequency dependent effective Coulomb matrix
elements. For graphene and graphite,  the effective Coulomb interaction is significantly
frequency dependent above $\omega\gtrsim 5$\,eV (Fig. \ref{fig:U_omega}).
Within the energy range of the Dirac spectrum, however, ($\sim 2$\,eV) the Coulomb
interaction can be well considered in the static limit.

In conclusion, the strength of Coulomb interactions in graphene
and graphite is accurately determined by first-principles
calculations. The local Coulomb interaction in graphene is
$U^{A/B}_{00}=9.3{\,\rm eV}\approx 3.3t$, which is very close to the critical value $U_{\rm sl}=3.5t$ for the transition to a gapped spin liquid. By straining graphene, the system can be driven across this critical value. Moreover, we find large non-local Coulomb interactions (e.g. $U_{01}=5.5\,\text{eV}\approx 2.0t$). By means of a dielectric substrate below graphene the screening of the long range tails of the Coulomb interaction can be tuned, while the local Coulomb interaction terms are expected to be much less affected by the dielectric environment. Hence, also the ratio of local to non-local Coulomb interactions can be tuned. It remains to be seen which additional many body instabilities might be triggered in this way or to which extent the conducting state of free standing graphene can be stabilized by non-local Coulomb terms. This issue deserves future attention. Very likely, our finding of large non-local Coulomb interaction $U_{01}$ in graphene generalizes to other two-dimensional materials.
In narrow impurity bands or edge states of graphene, the Coulomb
interaction might in any case present the dominating energy scale and, thus,
trigger many body instabilities including magnetism.

Financial support by the Deutsche Forschungsgemeinschaft through FOR-912, FOR-1346,
SFB 668, SPP 1145, SPP 1459 and FOM (The Netherlands) is acknowledged. TOW
is grateful to FZ J\"ulich for hospitality during the visit, when
parts of this work were conceived. One of us, SB, thanks Achim Rosch for fruitful discussions.

\bibliography{Umatrix_graphene2}

\end{document}